\documentclass[withtimes]{easychair}

\usepackage{doc}
\usepackage{acronym}
\usepackage{tikz}
\usepackage{pgfplots}
\pgfplotsset{compat = 1.3}
\usetikzlibrary{arrows}
\usetikzlibrary{positioning}

\newcommand{\fref}[1]{Figure~\ref{#1}}

\newcommand{\cref}[1]{Chapter~\ref{#1}}

\usepackage[sorting=none,firstinits=true,backend=bibtex,url=false]{biblatex}
\title{Intermittent Opportunistic Routing Components for the INET Framework
\footnote{This work was partially supported by the UK EPSRC under EP/P010164/1. \newline The source code is available at: \url{https://github.com/UoS-EEC/INET-opportunistic-routing}}}

\author{
Edward Longman, Mohammed El-Hajjar, Geoff V. Merrett
}

\institute{
  Electronics and Computer Science,
  University of Southampton, United Kingdom\\
  \email{\{el7g15,meh,gvm\}@ecs.soton.ac.uk}
 }

\authorrunning{Longman, El-Hajjar and Merrett}

\titlerunning{INET Intermittent Opportunistic Routing}

\newacro{OR}{opportunistic routing}
\newacro{IoT}{Internet of Things}
\newacro{WSN}{wireless sensor network}
\newacro{MAC}{medium access control}
\newacro{RPL}{routing protocol for low power lossy networks}
\newacro{ORW}{opportunistic routing for \ac{WSN}}
\newacro{ORPL}{opportunistic \ac{RPL}}
\newacro{DLL}{data link layer}
\newacro{NET}{network layer}
\usepackage{graphicx}

\begin{document}

\maketitle

\begin{abstract}

Intermittently-powered \acp{WSN} use energy harvesting and small energy storage to remove the need for battery replacement and to extend the operational lifetime. However, an intermittently-powered forwarder regularly turns on or off, which requires alternative networking solutions.
\Acf{OR} is a potential cross-layer solution for this novel application, but due to the interaction with the energy storage, the operation of these protocols is highly dynamic.
To compare protocols and components in like-for-like scenarios we propose module interfaces for  \acs{MAC}, routing and discovery protocols, that enable clear separation of concerns and good interchangeability.
We also suggest some candidates for each of the protocols based on our own implementation and research. \

\end{abstract}

%
%

\acresetall
\acused{MAC}
\acused{RPL}
\acused{DLL}
\acused{NET}
\section{Introduction}
\label{sect:introduction}

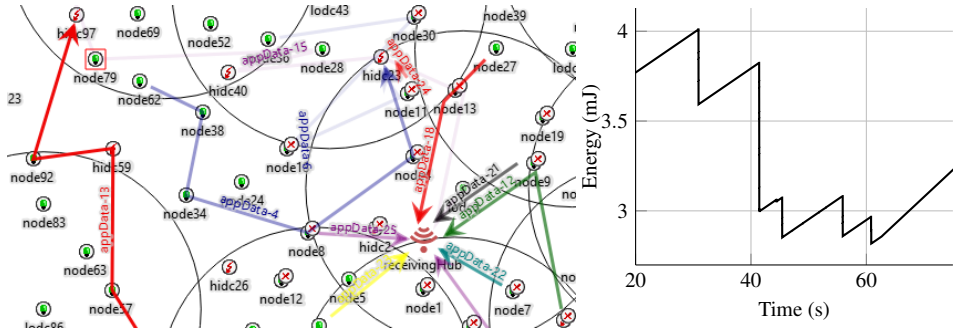
\begin{figure}
        \centering
        \includegraphics[width=0.52\textwidth]{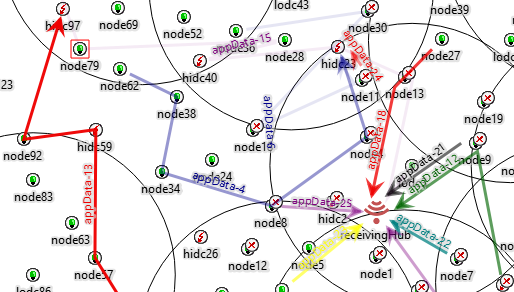}%
\begin{tikzpicture}
  \footnotesize
  \begin{axis}[
      width=0.4\textwidth, height=5cm,
      bar width=0.4,
      bar shift=0.2,
      axis x line*=bottom,
      axis y line*=left,
      grid=major,
      xlabel={Time (s)},
      ylabel={Energy (mJ)},
      xmax=75,xmin=20,
      ylabel shift = -0.7em
    ]
    \addplot[mark=none, thick] table [col sep=comma, x=tFrom10000, y={lodc86milliEnergy}]
      {lodc86PlottingIntermittencyOperation.csv};
  \end{axis}
  \end{tikzpicture}
        \caption{Left: \acs{OR} multi-point routing showing route taken through intermittently-powered nodes, where nodes marked with cross are off. Range marked by circles around some nodes. Right: Energy profile of harvest-transmit-cycle, where communication causes shutdowns to replenish the energy.}\label{fig:intermittent_nw_example}
\end{figure}

\Acp{WSN} are typically powered by batteries, which can be topped up by energy harvesting, but once empty they cannot be restarted. In future \ac{IoT} network solutions that do not require batteries but still with long lifetimes are required.
In intermittently-powered networks, nodes use energy harvesting and a small energy store~\cite{low_hanging_fruit}, but to reduce footprint and cost, the device is powered sporadically, as shown in \fref{fig:intermittent_nw_example}. However, current communications techniques for such devices rely on high capability forwarding nodes being within range, or being visited by mobile nodes~\cite{low_hanging_fruit,udugama_OPS}.
Instead, we focus on multi-hop  communication between intermittently-powered devices~\cite{geissdoerfer_find,rajib_predictive_transmission}, where communication events are limited by the energy storage and nodes are inherently only intermittently-connected.

\Ac{OR} is a viable networking paradigm for intermittently-powered networks~\cite{low_hanging_fruit}, as in \fref{fig:intermittent_nw_example}, that uses flexible forwarding to reach a specified network destination. Flexible forwarding is extremely important in intermittently-connected networks, where a complete route, from the source to the destination, cannot be established.
Finding which \ac{OR} method to use or improve depends on factors such as the node range, inter-node contact time, node energy supply and anticipated network throughput.

\section{Networking of Intermittently Powered Devices}
\label{sect:typesetting}

Intermittently-powered devices frequently and unpredictably shutdown, and then restart when there is sufficient stored energy.
This dynamically affects the operation of the network because devices experience varying loads, interleaved duty cycle~\cite{geissdoerfer_find} or shut down whilst waiting to forward data. Consequently there is a need for  simulation with the power consumption in the loop, and across multiple nodes.

Existing \ac{OR} implementations for OMNeT do not consider  power consumption~\cite{udugama_OPS} and are instead designed for  higher capability mobile   device scenarios.
However, \ac{WSN} nodes have limited energy available to communicate position, availability and routing sets. Additionally, data copying, bundle and broadcast methods exhaust network energy supplies, since data is duplicated and transmitted more than necessary.

Therefore, we propose \ac{OR} implementations for  INET that harness the existing energy storage, traffic  and advanced radio models. Since most opportunistic and cooperative routing protocols share common features and requirements this motivates our proposed interface.

\section{Common Opportunistic Behaviour and Requirements}

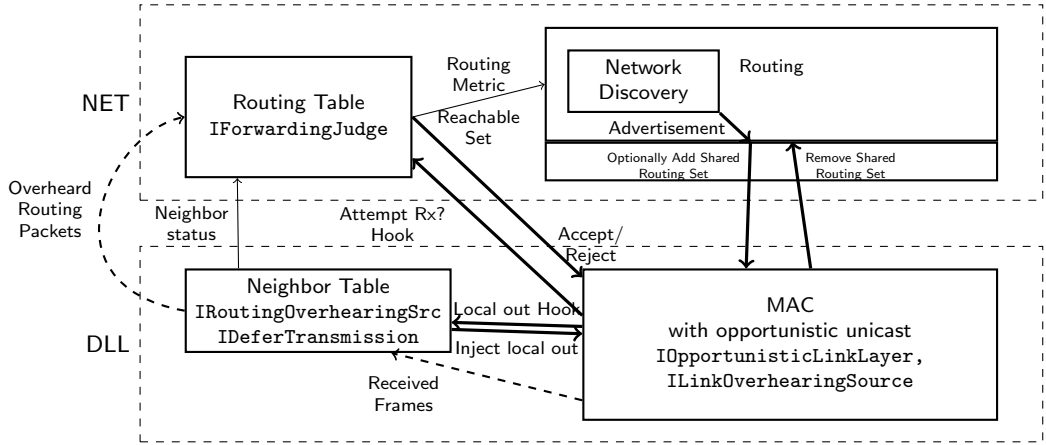
\begin{figure}
\centering
\begin{tikzpicture}[align=center,font={\scriptsize\sffamily}]
\tikzstyle{LayerGroup} = [draw=black,dashed,thin,rectangle,minimum width=120mm,minimum                 height=26mm,label=]
\tikzstyle{Signal} = [->,dashed, thick]
\tikzstyle{Component} = [draw=black,thick,rectangle,
                minimum width=60mm,minimum height=8mm,font={\footnotesize\sffamily}]
        \node[label=left:{\small\sffamily DLL}, LayerGroup](DLL){};

        \node[label=left:{\small\sffamily NET},above = 6mm of DLL,LayerGroup](NET){};
        \node[above right = 3mm and 6mm of NET.south west,
                anchor=south west,Component,
                minimum width=30mm, minimum height=16mm]
                (Routing_table){Routing Table\\\texttt{IForwardingJudge}};
        \node[below left =  3mm and 6mm of NET.north east,
                anchor=north east,Component,minimum height=15mm]
                (Routing){};
        \node[below = 3mm of Routing.north, anchor=north]
                (Routing_label){Routing};
        \node[below right = 3mm and 3mm of Routing.north west,
                anchor=north west,Component,minimum width=20mm]
                (Discovery){Network\\ Discovery};
        \node[below = 0 of Routing.south,
                anchor=north,Component,minimum height=5mm]
                (Announcement){};
        \draw(Discovery.south east) --
                node[left]{Advertisement} (Routing.250)[->, very thick];
        \node[below right =  3mm and 6mm of DLL.north west,
                anchor=north west,Component,minimum width=30mm]
                (Neighbours){Neighbor Table                \\\texttt{IRoutingOverhearingSrc}\\ \texttt{IDeferTransmission}  };
        \node[below left = 3mm and 6mm of DLL.north east, 
                anchor=north east,Component,minimum height=20mm, minimum width = 55mm]
                (MAC){MAC\\ with opportunistic unicast\\ \texttt{IOpportunisticLinkLayer,}\\  \texttt{ILinkOverhearingSource}
};
        \draw (MAC.195) --  
                node[below left, pos=0.7]{Received\\ Frames} 
                (Neighbours.330)[Signal];
        \draw (Neighbours.180) .. controls +(170:2cm) 
                and +(190:1cm) .. node[left, pos=0.5]{Overheard\\Routing\\Packets} (Routing_table.west)[Signal];
        \draw (MAC.172) -- 
                node[below left,pos=0.75]{Attempt Rx?\\ Hook}
                (Routing_table.340)[->, very thick];
        \draw (Routing_table.east) --  
                node[right,pos=0.8]{Accept/\\Reject} 
                (MAC.162)[->, very thick];
        \draw (MAC.175) -- 
                node[above]{Local out Hook}
                (Neighbours.355)[->, very thick];
        \draw (Neighbours.352) -- 
                node[below]{Inject local out}
                (MAC.177)[->, very thick];
        \draw (Neighbours.152) --  
                node[left]{Neighbor\\status} 
                (Routing_table.225)[->];
        \draw (Routing_table.east) --  
                node[below]{Reachable\\Set} 
                node[above]{Routing\\Metric} 
                (Routing.west)[->];
        \draw (MAC.75) -- 
                node[right,pos=0.8,font={\tiny\sffamily}]
                        {Remove Shared \\ Routing Set } 
                        (Routing.290)[->, very thick];
        \draw (Routing.250) -- 
                node[left,pos=0.2,font={\tiny\sffamily}]
                        {Optionally Add Shared \\ Routing Set }  
                        (MAC.120)[->, very thick];
\end{tikzpicture}
\vspace{-1em}
\caption{\Ac{OR} cross layer interfaces for network discovery, packet deferral, overhearing and acceptance.         }\label{fig:Oppostack_generic}
\vspace{-1em}
\end{figure}

\Ac{OR} operates across the \ac{DLL} and \ac{NET} layers, harnessing a broadcast link to reach several potential forwarding nodes. When the routing layer has data to send, it is passed down to the \ac{MAC} layer with attached forwarder selection criteria to select permitted forwarding nodes.
When the packet is broadcast, \emph{receiving nodes} must decide whether they meet the criteria for forwarding, and attempt reception of the packet using a cross layer query, as shown in \fref{fig:Oppostack_generic}. Following acceptance, there may be further \ac{MAC} layer contention for the packet before it is sent to the routing layer.

The first proposal is the interface, \texttt{IOpportunisticLinkLayer}, that should be implemented by any \ac{MAC} protocol that will check the acceptance criteria using a hook or message before contending to become the forwarder of a received packet.
Any response from the routing layer is timing dependent and must therefore occur before the receive window elapses. This is shown in \fref{fig:Oppostack_generic}, by the direct connection between the Routing Table and \ac{MAC}.
The corresponding interface is implemented by the routing table, \texttt{IForwardingJudge}.


Regardless of the \ac{OR} mechanism, the routing layer must build up a model of the network for direct addressing, a preferred forwarder list and calculating a progress metric. It also must store and disseminate information about the reachable set of neighbors.

The second proposal is a signal-subscribe interface, \texttt{ILinkOverhearingSource}, 
that emits signals for each type of encounter, for example whether it is coincidental or expected in response to an acknowledgment.
The receiver of these emitted signals must be able to understand \ac{MAC} layer datagrams and should then implement \texttt{IRoutingOverhearingSource}%
. This module could predict neighbor availability and implement \texttt{IDeferTransmission} to improve reception probability~\cite{rajib_predictive_transmission}.


Following from the neighbor overhearing, we also propose routing set middleware, for which we use the \texttt{TlvOption} specification as a starting point. This piggybacks the routing set into a proportion of packets, and hence reduces the need for extra advertisement packets. The routing set overhearing is handled by listening to  \texttt{IRoutingOverhearingSource}%
.

\subsection{Implementation of existing protocols}

The structure
of the above proposals comes from our experience implementing the \ac{ORPL} protocol~\cite{duquennoy_ORPL},  and studying techniques to improve neighbor discovery and packet throughput.
For example, our naive neighbor discovery protocol could be replaced by Find~\cite{geissdoerfer_find}, a discovery protocol specifically tuned for intermittent devices. Likewise, neighbor prediction could benefit from an advanced predictor tailored for intermittent networks~\cite{rajib_predictive_transmission}, which would implement \texttt{IDeferTransmission}.

The opportunistic operation of \ac{ORPL} can be seen in \fref{fig:intermittent_nw_example} demonstrating the variety of routes taken, where each next-hop decision depends on the instantaneous availability of forwarding nodes.
This implementation makes use of the proposed interfaces to encourage comparisons to new protocols.

\subsection{Working Implementation using Opportunistic Interfaces}

The  implementation is based on \acf{ORW}\cite{ghadimi_opportunistic_2014} and implements the described interfaces. \ac{ORW} uses neighbor encounter detection to calculate and expected cost (in terms of node duty cycle)  to the sink, termed EDC. Our implementation implements both the neighbor table, \ac{MAC} forwarder negotiation protocol and the network layer routing protocol. While  the implementation does not yet exactly match our proposed interfaces, we are working towards this. The source code is available at
\url{https://github.com/UoS-EEC/INET-opportunistic-routing}, and is briefly discussed here.

The \ac{MAC} protocol \texttt{ORWMac} implements the \texttt{IOpportunisticLinkLayer} interface, where the accept reject interface is inherited from \texttt{inet::NetFilterBase}. However, the preferred interface could use a query out and  accept in gates passing the received packet temporarily to the routing layer.
The signals emitted when the link layer overhears an incoming transmission are categorized depending on if it was expected, for example an acknowledgement response to a transmission,
or coincidental, for example an initial packet reception or advertisement
packets.
Also, a signal is emitted at the end of a period where  packets are  expected, regardless of whether they are actually received.

\begin{figure}
\centering
\includegraphics[width=0.6\textwidth]{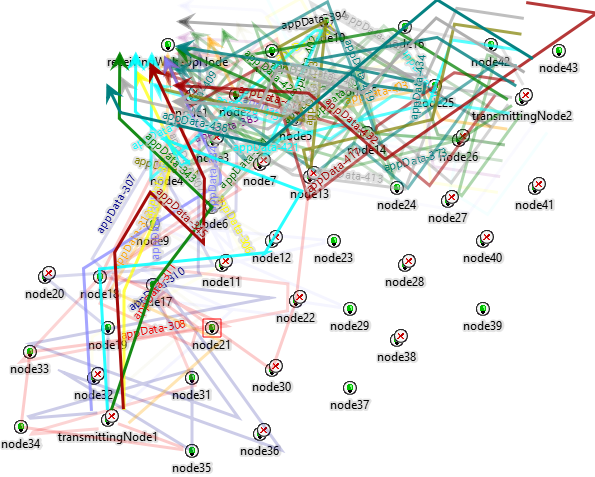}
\vspace{-1em}
\caption{Operation of Many hop \ac{ORW} demonstration with intermittently-powered forwarding nodes, showing variety of routes taken from two transmitting nodes to single destination.}
\label{fig:multihop_demo}
\vspace{-1em}
\end{figure}

The routing table \texttt{ORWRoutingTable} provides the \texttt{calculateUpwardsCost(L3Address dest) }interface, for the \texttt{ORWRouting} protocol to tag outgoing packets. For incoming packets it implements the other half of the forwarding request interface with \texttt{inet::NetfilterBase::HookBase} and likewise the preferred interface could use message gates implementing \texttt{IForwardingJudge}.
The advantage of using the hooks is that the containing \texttt{NetworkLayerNodeBase} and \texttt{IWirelessInterface} do not need to be modified. A demonstration of upward routing  \ac{ORW} can be seen in \fref{fig:multihop_demo}.

We extend  the \texttt{ORWRouting} class
to enable downward routing implementing \ac{ORPL}~\cite{duquennoy_ORPL} in \texttt{ORPLRouting} and \texttt{ORPLRoutingTable}.
This allows any node in the network to be reachable from any other node and requires occasional routing set sharing.
The routing set is added to outgoing packets after routing has happened and removed before routing layer acceptance by \texttt{ORPLRouting},
the \texttt{ORPLRoutingTable} listens to \texttt{ILinkOverhearingSource}  for updating routing set information.

Since there is not currently a  distinction between the neighbor  and  routing table and because deferred transmissions are not implemented, \texttt{IRoutingOverhearingSource} and \texttt{IDeferTransmission}, are currently unused, but are useful elements that can be used to improve the \ac{OR} implementation.

\section{Conclusion}

\Ac{OR} can  route information in intermittently-powered \acp{WSN} and to characterize their performance, in consideration of the power consumption, we have developed interfaces to model them.
There are several techniques that purport to improve certain aspects such as neighbor availability prediction, or the routing itself. The proposed interfaces between modules allows for innovation and optimization of these narrow aspects of intermittent networking to be tested in a whole system, as well as against each other.
Our implementation enables testing of opportunistic protocols intermittently-powered networks to explore suitable solutions and parametric optimisation.

\label{sect:bib}
\footnotesize
\printbibliography

\end{document}